\documentclass[dvips]{elsart}

\usepackage{amsfonts,amsmath,amssymb}
\usepackage{dcolumn}
\usepackage{graphicx}
\usepackage[numbers,sort&compress]{natbib}
\usepackage{hypernat,hyperref}

\newcommand{\onlinecite}[1]{\cite{#1}}

\begin{document}

\begin{frontmatter}
   \title{Precise dipole moment and quadrupole coupling constants of benzonitrile}%

   \author[fhi]{Kirstin Wohlfart}, %
   \author[fhi]{Melanie Schnell}, %
   \author[han]{Jens-Uwe Grabow}, and %
   \author[fhi]{Jochen K\"upper\corauthref{correspond}} %
   \corauth[correspond]{Author to whom correspondence should be addressed. Electronic mail:
      jochen@fhi-berlin.mpg.de}%
   
   \address[fhi]{Fritz-Haber-Institut der Max-Planck-Gesellschaft, Faradayweg 4--6, 14195 Berlin,
      Germany}%
   \address[han]{Gottfried-Wilhelm-Leibniz-Universit\"at, Institut f\"ur Physikalische Chemie und
      Elektrochemie, Callinstra{\ss}e 3a, 30167 Hannover, Germany}%

   \date{\today}%

   \begin{keyword}
      benzonitrile; microwave spectroscopy; rotational spectroscopy; dipole moment; Stark effect;
      nuclear quadrupole coupling; supersonic-jet; cold molecules%
      \PACS{33.20.Bx; 33.55.Be; 33.15.-e}
   \end{keyword}

\end{frontmatter}

Cyanobenzenes have recently been suggested as good candidates for the detection of aromatic
molecules in interstellar space \cite{Woods:ApJ574:L167, Woods:IAUSymp2003:279,
   Herbst:privcomm:2007}, partly due to their large dipole moments. The simplest molecule from this
class is benzonitrile (C$_6$H$_5$CN), which might be detectable, for example, in the proto-planetary
nebula CRL~618~\cite{Woods:ApJ574:L167, Woods:IAUSymp2003:279}. These predictions are based on an
assumed abundance of $10^{-6}$ and a dipole moment of 4.3\,D~\cite{Woods:ApJ574:L167}, and they are
related to the detection of similar molecular species (benzene (C$_6$H$_6$), diacetylene
(C$_4$H$_2$), and triacetylene (C$_6$H$_2$)) in CRL~618~\cite{Cernicharo:ApJ546:L123}. Moreover, due
to its large dipole moment and therefore its large Stark-effect-to-mass ratio benzonitrile is a
prime candidate for the electrostatic deceleration and trapping of large molecules using the
Alternate Gradient focusing principle \cite{Bethlem:JPB39:R263}.

The value of the benzonitrile dipole moment, however, is still a matter of debate. Due to the
$C_{2v}$ symmetry only a non-zero $\mu_{\alpha}$ component of the dipole moment can exist. To our
knowledge, only two independent and incompatible values of the dipole moment of benzonitrile in its
electronic ground state have been published: From microwave Stark effect measurements a dipole
moment of 4.14(5)\,D was deduced~\cite{Lide:JCP22:1577} and a similar value based on the same
measurements (4.18(8)\,D) is listed in the CRC Handbook of Chemistry and
Physics~\cite{CRC:HandbookChemPhys71}. Recently, a different value of 4.48(1)\,D was determined from
Stark shift measurements of the rotationally resolved laser-induced-fluorescence spectrum
\cite{Borst:CPL350:485}. Therefore, we set out for a precise experimental determination of the
dipole moment using Fourier-transform microwave spectroscopy (FTMW) in homogeneous electric fields.


The experimental setup of the Hannover COBRA-FTMW-spectrometer has been described in detail
elsewhere~\cite{Grabow:RSI67:4072, Schnell:RSI75:2111}. In brief, benzonitrile ($\geq99$\,\% purity)
was purchased from Fluka and used without further purification. The sample was co-expanded in
3.5~bar of Ne at a temperature of 300\,K through a pulsed nozzle (General Valve Series 9) with a
0.8~mm orifice. The supersonic expansion was pulsed coaxially into the microwave
resonator~\cite{Grabow:ZNatA45:1043}, which was specially developed to provide high sensitivity and
resolution at low frequencies down to 2\,GHz~\cite{Grabow:Habilitation:2004}. Some of the lowest
rotational transitions of benzonitrile in the range of 2.8--10.9\,GHz were recorded with a linewidth
(FWHM) of 2.5\,kHz and a frequency accuracy of 500~Hz. Homogeneous electric fields for the Stark
shift measurements were provided by the Coaxially Aligned Electrodes for Stark effect Applied in
Resonators (CAESAR) setup~\cite{Schnell:RSI75:2111} where electric field strengths up to 203\,V/cm
could be reached. The electric field was calibrated using the $J=1\leftarrow{}0$ transition of
OC$^{36}$S (0.02\,\% natural abundance) and a documented dipole moment of
0.71519(3)\,D~\cite{Reinartz:CPL24:346}, see reference~\onlinecite{Filsinger:MW-3AP:inprep} for
details.


The measured field-free microwave transitions and their estimated experimental uncertainties are
listed in table \ref{tab:lines}.
\begin{table}
   \centering\scriptsize
   \begin{tabular}{c@{ $\leftarrow$ }cc@{ $\leftarrow$ }cccc}
      \hline\hline
      $J'_{K_a'\,K_c'}$ & $J''_{K_a''\,K_c''}$ & $F'$ & $F''$ & frequency & uncertainty & obs.$-$calc. \\
      \multicolumn{4}{c}{} & (MHz) & (MHz) & (MHz) \\
      \hline
      $1_{0\,1}$ & $0_{0\,0}$ & 1 & 1 &  2760.22090 & 0.00050 & 0.00030 \\
      $1_{0\,1}$ & $0_{0\,0}$ & 2 & 1 &  2761.49318 & 0.00150 & 0.00154 \\
      $1_{0\,1}$ & $0_{0\,0}$ & 0 & 1 &  2763.39853 & 0.00050 &-0.00011 \\
      $2_{0\,2}$ & $1_{0\,1}$ & 1 & 0 &  5502.12082 & 0.00100 &-0.00095 \\
      $2_{0\,2}$ & $1_{0\,1}$ & 1 & 1 &  5505.30033 & 0.00100 & 0.00052 \\
      $2_{0\,2}$ & $1_{0\,1}$ & 3 & 2 &  5503.27585 & 0.00100 & 0.00009 \\
      $2_{0\,2}$ & $1_{0\,1}$ & 2 & 1 &  5503.19315 & 0.00100 & 0.00092 \\
      $2_{0\,2}$ & $1_{0\,1}$ & 1 & 2 &  5504.02800 & 0.00100 &-0.00077 \\
      $2_{0\,2}$ & $1_{0\,1}$ & 2 & 2 &  5501.92073 & 0.00100 &-0.00046 \\
      $2_{0\,2}$ & $1_{0\,1}$ & 3 & 2 &  5503.27579 & 0.00100 & 0.00003 \\
      $2_{1\,1}$ & $1_{1\,0}$ & 1 & 0 &  5856.56986 & 0.00100 &-0.00059 \\
      $2_{1\,1}$ & $1_{1\,0}$ & 1 & 1 &  5855.10867 & 0.00150 &-0.00070 \\
      $2_{1\,1}$ & $1_{1\,0}$ & 2 & 1 &  5853.96436 & 0.00100 &-0.00029 \\
      $2_{1\,1}$ & $1_{1\,0}$ & 1 & 2 &  5855.69296 & 0.00100 &-0.00073 \\
      $2_{1\,1}$ & $1_{1\,0}$ & 2 & 2 &  5854.54929 & 0.00100 & 0.00032 \\
      $2_{1\,1}$ & $1_{1\,0}$ & 3 & 2 &  5855.28522 & 0.00100 & 0.00073 \\
      $2_{1\,2}$ & $1_{1\,1}$ & 1 & 0 &  5191.71753 & 0.00050 & 0.00000 \\
      $2_{1\,2}$ & $1_{1\,1}$ & 1 & 1 &  5190.00134 & 0.00050 &-0.00017 \\
      $2_{1\,2}$ & $1_{1\,1}$ & 2 & 1 &  5189.02665 & 0.00150 & 0.00008 \\
      $2_{1\,2}$ & $1_{1\,1}$ & 1 & 2 &  5190.68619 & 0.00150 & 0.00170 \\
      $2_{1\,2}$ & $1_{1\,1}$ & 3 & 2 &  5190.33981 & 0.00100 & 0.00062 \\
      $3_{0\,3}$ & $2_{0\,2}$ & 4 & 3 &  8206.82990 & 0.00100 & 0.00066 \\
      $3_{0\,3}$ & $2_{0\,2}$ & 3&  2 &  8206.79124 & 0.00100 & 0.00102 \\
      $3_{0\,3}$ & $2_{0\,2}$ & 2 & 1 &  8206.56481 & 0.00100 & 0.00051 \\
      $4_{0\,4}$ & $3_{0\,3}$ & 5 & 4 & 10855.26125 & 0.00150 & 0.00071 \\
      $4_{0\,4}$ & $3_{0\,3}$ & 4 & 3 & 10855.24427 & 0.00150 & 0.00092 \\
      $4_{0\,4}$ & $3_{0\,3}$ & 3 & 2 & 10855.13590 & 0.00100 & 0.00043 \\
      $4_{0\,4}$ & $3_{0\,3}$ & 3 & 3 & 10857.01706 & 0.00050 &-0.00008 \\
      $4_{0\,4}$ & $3_{0\,3}$ & 3 & 4 & 10855.62177 & 0.00200 &-0.00178 \\
      $4_{0\,4}$ & $3_{0\,3}$ & 4 & 4 & 10853.84943 & 0.00050 &-0.00032 \\
      \hline\hline
   \end{tabular}
   \caption{Measured hyperfine-resolved field-free transition frequencies, estimated experimental
      uncertainties, and fit residuals. See text for details.}
   \label{tab:lines}
\end{table}
Whereas the frequencies for the $J=1\leftarrow0$ and $2\leftarrow1$ hyperfine transitions are
reported for the first time the accuracy for the $J=3\leftarrow2$ and $4\leftarrow3$ transitions
have been considerably improved with respect to previous measurements~\cite{Fliege:ZNatA36:1124} and
additional hyperfine-splittings could be resolved.

We have simultaneously fit the newly measured lines together with previously published microwave
transitions~\cite{Fliege:ZNatA36:1124, Vormann:ZNatA43:283} using the QStark program
package~\cite{Kisiel:CPL325:523, Kisiel:JPC104:6970}. A weighted analysis of all these lines was
performed assuming for the transition frequencies from Fliege et al.~\cite{Fliege:ZNatA36:1124}
measurement accuracies of 20\,kHz and for the transition frequencies from Vormann et
al.~\cite{Vormann:ZNatA43:283} measurement accuracies of 5\,kHz. The resulting rotational constants,
centrifugal distortion constants, and nitrogen nuclear quadrupole coupling constants of benzonitrile
are given in table \ref{tab:rotcon}.
\begin{table}
   \centering
   \begin{tabular}{lc}
      \hline\hline
      $A$ (MHz)                   & 5655.2654\,(72)   \\
      $B$ (MHz)                   & 1546.875864\,(66) \\
      $C$ (MHz)                   & 1214.40399\,(10)  \\
      $\chi_{aa}$ (MHz)           &   -4.23738\,(36)  \\
      $\chi_{bb}-\chi_{cc}$ (MHz) &    0.3397\,(10)   \\
      $\chi_{bb}$ (MHz)           &    2.2886\,(11)   \\
      $\chi_{cc}$ (MHz)           &    1.9488\,(11)   \\
      $\Delta_J$ (kHz)            &    0.0456\,(15)   \\
      $\Delta_ {JK}$ (kHz)        &    0.9381\,(56)   \\
      $\Delta_K$ (kHz)            &    0.50(38)       \\
      $\delta_J$ (kHz)            &    0.01095\,(41)  \\
      $\delta_K$ (kHz)            &    0.628\,(53)    \\
      $\Delta{I}$ (u\AA$^2$)      &    0.0801\,(12)   \\
      number of measurements      &   93              \\
      $\sigma$ (MHz)              &  0.00524          \\
      $\hat{\sigma}$              &  0.675            \\
      \hline
      $\mu_a$ (D)                 &    4.5152\,(68)   \\
      number of measurements      &   78              \\
      $\sigma$ (MHz)              &  0.00228          \\
      $\hat{\sigma}$              &  0.709            \\
      \hline\hline
   \end{tabular}
   \caption{Upper part: Rotational constants, $^{14}$N quadruple coupling constants, centrifugal
      distortion constants, inertial defect, the number of hyperfine-resolved components included
      in the fit, overall standard deviation, and weighted standard deviation from the fit of the
      field-free lines of benzonitrile. Lower part: Dipole moment, number of measurements at
      different electric field strengths, overall standard deviation, and weighted standard
      deviation from the fit of the Stark-shifts at various electric field strengths. See text for
      details.}
   \label{tab:rotcon}
\end{table}
The values of the rotational constants are in good agreement with literature
values~\cite{Fliege:ZNatA36:1124, Vormann:ZNatA43:283, Wlodarczak:JMolSpec134:297}. The nitrogen
nuclear quadrupole coupling constants agree with those determined by the Kiel group using
perturbation theory~\cite{Fliege:ZNatA36:1124, Vormann:ZNatA43:283} within the published error
limits, and, moreover, the accuracy of these nuclear quadrupole coupling constants could be improved
considerably in our study. The centrifugal distortion constants agree within error limits with those
of large-$J$ microwave transition measurements~\cite{Wlodarczak:JMolSpec134:297}. The resulting
value of the inertial defect $\Delta I=0.0801\,(12)$\,u\AA$^2$ is small and confirms the planarity
of benzonitrile.

In order to determine the dipole moment of benzonitrile, Stark effect measurements in homogeneous
electric fields were performed on the
$J'_{K_a'\,K_c'}F'\leftarrow{}J''_{K_a''\,K_c''}F''=1_{0\,1}1\leftarrow0_{0\,0}0$,
$1_{0\,1}1\leftarrow0_{0\,0}1$, and $4_{0\,4}3\leftarrow3_{0\,3}3$ transitions. For our deceleration
experiments we are especially interested in the Stark effect of the absolute rovibronic ground state
of benzonitrile. Moreover, that $0_{00}$ state is the most polar rotational state for all molecules,
and, therefore, exhibits the largest Stark shift in strong electric
fields~\cite{Bethlem:JPB39:R263}. Additionally, the selected transition had good intensities. The
applied high voltages were calibrated
and electric field strengths calibration measurements for OCS were done, as described in
reference~\onlinecite{Filsinger:MW-3AP:inprep}. From these measurements an error estimate for the
electric field strengths was specified. For electric fields up to 203\,V/cm the frequencies of the
transitions in the electric field could be determined with uncertainties between 0.5~kHz--8\,kHz,
depending on the $J$-complexity of the Stark-split and -shifted spectra and the electric field
strength. The dipole moment and its error estimate are given in table~\ref{tab:rotcon}. This value
is in agreement ($3\sigma$) with the value from Stark-shift measurements of the rotationally
resolved laser-induced-fluorescence spectrum~\cite{Borst:CPL350:485}, but differs considerably from
previous microwave measurements~\cite{Lide:JCP22:1577, CRC:HandbookChemPhys71}.\footnote{It should
   be noted, that in the original publication the author himself states that ``attempts to study the
   Stark effect of benzonitrile encountered a number of difficulties.'' See
   reference~\onlinecite{Lide:JCP22:1577} for details.}

In conclusion, we have simultaneously determined the rotational constants, nitrogen nuclear
quadrupole coupling constants, and centrifugal distortion constants of benzonitrile from Fourier
transform microwave spectroscopy in a supersonic jet. The value of the dipole moment of benzonitrile
is determined precisely to $\mu_a=4.5152\,(68)$~D from Stark-shift measurements. This should settle
the issue regarding the value of the dipole moment of benzonitrile.

\begin{center}
   \textbf{Acknowledgments}
\end{center}
We would like to thank Gerard Meijer for his continuous support and Zbigniew Kisiel for updating
QStark to allow weighted fits. Financial support from the \emph{Land Niedersachsen} and the
\emph{Deutsche Forschungsgemeinschaft}, also within the priority program 1116 ``Interactions in
ultracold gases'', is acknowledged.

\bibliographystyle{apsrev-nourl}
\bibliography{string,mp}

\begin{thebibliography}{20}
\expandafter\ifx\csname natexlab\endcsname\relax\def\natexlab#1{#1}\fi
\expandafter\ifx\csname bibnamefont\endcsname\relax
  \def\bibnamefont#1{#1}\fi
\expandafter\ifx\csname bibfnamefont\endcsname\relax
  \def\bibfnamefont#1{#1}\fi
\expandafter\ifx\csname citenamefont\endcsname\relax
  \def\citenamefont#1{#1}\fi
\expandafter\ifx\csname url\endcsname\relax
  \def\url#1{\texttt{#1}}\fi
\expandafter\ifx\csname urlprefix\endcsname\relax\def\urlprefix{URL }\fi
\providecommand{\bibinfo}[2]{#2}
\providecommand{\eprint}[2][]{\url{#2}}

\bibitem[{\citenamefont{Woods et~al.}(2002)\citenamefont{Woods, Millar,
  Zijlstra, and Herbst}}]{Woods:ApJ574:L167}
\bibinfo{author}{\bibfnamefont{P.~M.} \bibnamefont{Woods}},
  \bibinfo{author}{\bibfnamefont{T.~J.} \bibnamefont{Millar}},
  \bibinfo{author}{\bibfnamefont{A.~A.} \bibnamefont{Zijlstra}},
  \bibnamefont{and} \bibinfo{author}{\bibfnamefont{E.}~\bibnamefont{Herbst}},
  \bibinfo{journal}{Astrophys. J.} \textbf{\bibinfo{volume}{574}},
  \bibinfo{pages}{L167} (\bibinfo{year}{2002}).

\bibitem[{\citenamefont{Woods et~al.}(2003)\citenamefont{Woods, Millar,
  Zijlstra, and Herbst}}]{Woods:IAUSymp2003:279}
\bibinfo{author}{\bibfnamefont{P.~M.} \bibnamefont{Woods}},
  \bibinfo{author}{\bibfnamefont{T.~J.} \bibnamefont{Millar}},
  \bibinfo{author}{\bibfnamefont{A.~A.} \bibnamefont{Zijlstra}},
  \bibnamefont{and} \bibinfo{author}{\bibfnamefont{E.}~\bibnamefont{Herbst}},
  \bibinfo{journal}{{IAU} SYMPOSIA: Planetary Nebulae: Their Evolution and Role
  in the Universe} pp. \bibinfo{pages}{279--280} (\bibinfo{year}{2003}).

\bibitem[{\citenamefont{Herbst}(2007)}]{Herbst:privcomm:2007}
\bibinfo{author}{\bibfnamefont{E.}~\bibnamefont{Herbst}},
  \bibinfo{howpublished}{private communication with J.-U. Grabow}
  (\bibinfo{year}{2007}).

\bibitem[{\citenamefont{Cernicharo et~al.}(2001)\citenamefont{Cernicharo,
  Heras, Tielens, Pardo, Herpin, Gu\'elin, and
  Waters}}]{Cernicharo:ApJ546:L123}
\bibinfo{author}{\bibfnamefont{J.}~\bibnamefont{Cernicharo}},
  \bibinfo{author}{\bibfnamefont{A.~M.} \bibnamefont{Heras}},
  \bibinfo{author}{\bibfnamefont{A.~G. G.~M.} \bibnamefont{Tielens}},
  \bibinfo{author}{\bibfnamefont{J.~R.} \bibnamefont{Pardo}},
  \bibinfo{author}{\bibfnamefont{F.}~\bibnamefont{Herpin}},
  \bibinfo{author}{\bibfnamefont{M.}~\bibnamefont{Gu\'elin}}, \bibnamefont{and}
  \bibinfo{author}{\bibfnamefont{L.~B. F.~M.} \bibnamefont{Waters}},
  \bibinfo{journal}{Astrophys. J.} \textbf{\bibinfo{volume}{546}},
  \bibinfo{pages}{L123} (\bibinfo{year}{2001}).

\bibitem[{\citenamefont{Bethlem et~al.}(2006)\citenamefont{Bethlem, Tarbutt,
  K\"upper, Carty, Wohlfart, Hinds, and Meijer}}]{Bethlem:JPB39:R263}
\bibinfo{author}{\bibfnamefont{H.~L.} \bibnamefont{Bethlem}},
  \bibinfo{author}{\bibfnamefont{M.~R.} \bibnamefont{Tarbutt}},
  \bibinfo{author}{\bibfnamefont{J.}~\bibnamefont{K\"upper}},
  \bibinfo{author}{\bibfnamefont{D.}~\bibnamefont{Carty}},
  \bibinfo{author}{\bibfnamefont{K.}~\bibnamefont{Wohlfart}},
  \bibinfo{author}{\bibfnamefont{E.~A.} \bibnamefont{Hinds}}, \bibnamefont{and}
  \bibinfo{author}{\bibfnamefont{G.}~\bibnamefont{Meijer}},
  \bibinfo{journal}{J. Phys. B} \textbf{\bibinfo{volume}{39}},
  \bibinfo{pages}{R263} (\bibinfo{year}{2006}).

\bibitem[{\citenamefont{Lide}(1954)}]{Lide:JCP22:1577}
\bibinfo{author}{\bibfnamefont{D.~R.} \bibnamefont{Lide}}, \bibinfo{journal}{J.
  Chem. Phys.} \textbf{\bibinfo{volume}{22}}, \bibinfo{pages}{1577}
  (\bibinfo{year}{1954}).

\bibitem[{\citenamefont{Lide}(1990)}]{CRC:HandbookChemPhys71}
\bibinfo{editor}{\bibfnamefont{D.~R.} \bibnamefont{Lide}}, ed.,
  \emph{\bibinfo{title}{CRC Handbook of Chemistry and Physics}},
  vol.~\bibinfo{volume}{71} (\bibinfo{publisher}{CRC Press},
  \bibinfo{address}{Boca Raton}, \bibinfo{year}{1990}).

\bibitem[{\citenamefont{Borst et~al.}(2001)\citenamefont{Borst, Korter, and
  Pratt}}]{Borst:CPL350:485}
\bibinfo{author}{\bibfnamefont{D.~R.} \bibnamefont{Borst}},
  \bibinfo{author}{\bibfnamefont{T.~M.} \bibnamefont{Korter}},
  \bibnamefont{and} \bibinfo{author}{\bibfnamefont{D.~W.} \bibnamefont{Pratt}},
  \bibinfo{journal}{Chem. Phys. Lett.} \textbf{\bibinfo{volume}{350}},
  \bibinfo{pages}{485} (\bibinfo{year}{2001}).

\bibitem[{\citenamefont{Grabow et~al.}(1996)\citenamefont{Grabow, Stahl, and
  Dreizler}}]{Grabow:RSI67:4072}
\bibinfo{author}{\bibfnamefont{J.-U.} \bibnamefont{Grabow}},
  \bibinfo{author}{\bibfnamefont{W.}~\bibnamefont{Stahl}}, \bibnamefont{and}
  \bibinfo{author}{\bibfnamefont{H.}~\bibnamefont{Dreizler}},
  \bibinfo{journal}{Rev. Sci. Instrum.} \textbf{\bibinfo{volume}{67}},
  \bibinfo{pages}{4072} (\bibinfo{year}{1996}).

\bibitem[{\citenamefont{Schnell et~al.}(2004)\citenamefont{Schnell, Banser, and
  Grabow}}]{Schnell:RSI75:2111}
\bibinfo{author}{\bibfnamefont{M.}~\bibnamefont{Schnell}},
  \bibinfo{author}{\bibfnamefont{D.}~\bibnamefont{Banser}}, \bibnamefont{and}
  \bibinfo{author}{\bibfnamefont{J.-U.} \bibnamefont{Grabow}},
  \bibinfo{journal}{Rev. Sci. Instrum.} \textbf{\bibinfo{volume}{75}},
  \bibinfo{pages}{2111} (\bibinfo{year}{2004}).

\bibitem[{\citenamefont{Grabow and Stahl}(1990)}]{Grabow:ZNatA45:1043}
\bibinfo{author}{\bibfnamefont{J.-U.} \bibnamefont{Grabow}} \bibnamefont{and}
  \bibinfo{author}{\bibfnamefont{W.}~\bibnamefont{Stahl}}, \bibinfo{journal}{Z.
  Naturforsch. A} \textbf{\bibinfo{volume}{45}}, \bibinfo{pages}{1043}
  (\bibinfo{year}{1990}).

\bibitem[{\citenamefont{Grabow}(2004)}]{Grabow:Habilitation:2004}
\bibinfo{author}{\bibfnamefont{J.-U.} \bibnamefont{Grabow}},
  \bibinfo{type}{Habilitationsschrift},
  \bibinfo{school}{Gottfried-Wilhelm-Leibniz-Universit\"at Hannover},
  \bibinfo{address}{Germany} (\bibinfo{year}{2004}).

\bibitem[{\citenamefont{Reinartz and Dymanus}(1974)}]{Reinartz:CPL24:346}
\bibinfo{author}{\bibfnamefont{J.~M. L.~J.} \bibnamefont{Reinartz}}
  \bibnamefont{and} \bibinfo{author}{\bibfnamefont{A.}~\bibnamefont{Dymanus}},
  \bibinfo{journal}{Chem. Phys. Lett.} \textbf{\bibinfo{volume}{24}},
  \bibinfo{pages}{346} (\bibinfo{year}{1974}).

\bibitem[{\citenamefont{Filsinger et~al.}(2007)\citenamefont{Filsinger,
  Wohlfart, Schnell, Grabow, and K\"upper}}]{Filsinger:MW-3AP:inprep}
\bibinfo{author}{\bibfnamefont{F.}~\bibnamefont{Filsinger}},
  \bibinfo{author}{\bibfnamefont{K.}~\bibnamefont{Wohlfart}},
  \bibinfo{author}{\bibfnamefont{M.}~\bibnamefont{Schnell}},
  \bibinfo{author}{\bibfnamefont{J.-U.} \bibnamefont{Grabow}},
  \bibnamefont{and} \bibinfo{author}{\bibfnamefont{J.}~\bibnamefont{K\"upper}},
  \bibinfo{journal}{Phys. Chem. Chem. Phys.}  (\bibinfo{year}{2007}),
  \bibinfo{note}{accepted, available as advance articles on the web: DOI:
  10.1039/b711888k}.

\bibitem[{\citenamefont{Fliege et~al.}(1981)\citenamefont{Fliege, Bestmann,
  Schwarz, and Dreizler}}]{Fliege:ZNatA36:1124}
\bibinfo{author}{\bibfnamefont{E.}~\bibnamefont{Fliege}},
  \bibinfo{author}{\bibfnamefont{G.}~\bibnamefont{Bestmann}},
  \bibinfo{author}{\bibfnamefont{R.}~\bibnamefont{Schwarz}}, \bibnamefont{and}
  \bibinfo{author}{\bibfnamefont{H.}~\bibnamefont{Dreizler}},
  \bibinfo{journal}{Z. Naturforsch. A} \textbf{\bibinfo{volume}{36}},
  \bibinfo{pages}{1124} (\bibinfo{year}{1981}).

\bibitem[{\citenamefont{Vormann et~al.}(1988)\citenamefont{Vormann, Andresen,
  Heinsing, and Dreizler}}]{Vormann:ZNatA43:283}
\bibinfo{author}{\bibfnamefont{K.}~\bibnamefont{Vormann}},
  \bibinfo{author}{\bibfnamefont{U.}~\bibnamefont{Andresen}},
  \bibinfo{author}{\bibfnamefont{N.}~\bibnamefont{Heinsing}}, \bibnamefont{and}
  \bibinfo{author}{\bibfnamefont{H.}~\bibnamefont{Dreizler}},
  \bibinfo{journal}{Z. Naturforsch. A} \textbf{\bibinfo{volume}{43}},
  \bibinfo{pages}{283} (\bibinfo{year}{1988}).

\bibitem[{\citenamefont{Kisiel et~al.}(2000{\natexlab{a}})\citenamefont{Kisiel,
  Kosarzewski, Pietrewicz, and Pszcz\'o{\l}kowski}}]{Kisiel:CPL325:523}
\bibinfo{author}{\bibfnamefont{Z.}~\bibnamefont{Kisiel}},
  \bibinfo{author}{\bibfnamefont{J.}~\bibnamefont{Kosarzewski}},
  \bibinfo{author}{\bibfnamefont{B.~A.} \bibnamefont{Pietrewicz}},
  \bibnamefont{and}
  \bibinfo{author}{\bibfnamefont{L.}~\bibnamefont{Pszcz\'o{\l}kowski}},
  \bibinfo{journal}{Chem. Phys. Lett.} \textbf{\bibinfo{volume}{325}},
  \bibinfo{pages}{523} (\bibinfo{year}{2000}{\natexlab{a}}).

\bibitem[{\citenamefont{Kisiel et~al.}(2000{\natexlab{b}})\citenamefont{Kisiel,
  Pietrewicz, Fowler, Legon, and Steiner}}]{Kisiel:JPC104:6970}
\bibinfo{author}{\bibfnamefont{Z.}~\bibnamefont{Kisiel}},
  \bibinfo{author}{\bibfnamefont{B.~A.} \bibnamefont{Pietrewicz}},
  \bibinfo{author}{\bibfnamefont{P.~W.} \bibnamefont{Fowler}},
  \bibinfo{author}{\bibfnamefont{A.~C.} \bibnamefont{Legon}}, \bibnamefont{and}
  \bibinfo{author}{\bibfnamefont{E.}~\bibnamefont{Steiner}},
  \bibinfo{journal}{J. Phys. Chem. A} \textbf{\bibinfo{volume}{104}},
  \bibinfo{pages}{6970} (\bibinfo{year}{2000}{\natexlab{b}}).

\bibitem[{\citenamefont{Unterberg et~al.}(2004)\citenamefont{Unterberg,
  Gerlach, Jansen, and Gerhards}}]{Unterberg:CP304:237}
\bibinfo{author}{\bibfnamefont{C.}~\bibnamefont{Unterberg}},
  \bibinfo{author}{\bibfnamefont{A.}~\bibnamefont{Gerlach}},
  \bibinfo{author}{\bibfnamefont{A.}~\bibnamefont{Jansen}}, \bibnamefont{and}
  \bibinfo{author}{\bibfnamefont{M.}~\bibnamefont{Gerhards}},
  \bibinfo{journal}{Chem. Phys.} \textbf{\bibinfo{volume}{304}},
  \bibinfo{pages}{237} (\bibinfo{year}{2004}).

\bibitem[{\citenamefont{Wlodarczak et~al.}(1989)\citenamefont{Wlodarczak,
  Burie, Demaison, Vormann, and Cs\'asz\'ar}}]{Wlodarczak:JMolSpec134:297}
\bibinfo{author}{\bibfnamefont{G.}~\bibnamefont{Wlodarczak}},
  \bibinfo{author}{\bibfnamefont{J.}~\bibnamefont{Burie}},
  \bibinfo{author}{\bibfnamefont{J.}~\bibnamefont{Demaison}},
  \bibinfo{author}{\bibfnamefont{K.}~\bibnamefont{Vormann}}, \bibnamefont{and}
  \bibinfo{author}{\bibfnamefont{A.~G.} \bibnamefont{Cs\'asz\'ar}},
  \bibinfo{journal}{J. Mol. Spec.} \textbf{\bibinfo{volume}{134}},
  \bibinfo{pages}{297} (\bibinfo{year}{1989}).

\end{thebibliography}

\end{document}